\documentclass[twocolumn,showpacs,preprintnumbers,aps,prd,a4paper,
superscriptaddress,nofootinbib,tightenlines,floats,floatfix]{revtex4}

\usepackage{bm}
\usepackage{epsfig}
\usepackage{citesort}
\usepackage{color}
 
\def\lsim{\mathrel{\mathop 
  {\hbox{\lower0.5ex\hbox{$\sim$}\kern-0.8em\lower-0.7ex\hbox{$<$}}}}} 
\def\gsim{\mathrel{\mathop
  {\hbox{\lower0.5ex\hbox{$\sim$}\kern-0.8em\lower-0.7ex\hbox{$>$}}}}}

\newcommand{\half}{{1\over2}} 
\newcommand{\nad}{n_{\rm ad}} 
\newcommand{\niso}{n_{\rm iso}}

\newcommand{\Ocdm}{\Omega_{\rm cdm}} 
\newcommand{\ocdm}{\omega_{\rm cdm}}

\newcommand{\oB}{\omega_{\rm b}} 
\newcommand{\Oa}{\Omega_{\rm a}}

\newcommand{\ma}{m_{a}} 
\newcommand{\na}{n_{a}} 
\newcommand{\be}{\begin{equation}} 
\newcommand{\ee}{\end{equation}} 
 
\newcommand{\fN}{\frac{f_a}{N}}
\newcommand{\fa}{f_a} 
\newcommand{\hinf}{H_{\rm inf}} 
\newcommand{\trh}{T_{\rm rh}} 
\newcommand{\MP}{M_{\rm P}} 
 
\newbox\pippobox
 
\begin{document}
 
\begin{flushright}
{\large \tt hep-ph/0606107, \tt LAPTH-1149/06}
\end{flushright}

\title{Isocurvature bounds on axions revisited} 
 
\author{Mar{\'\i}a Beltr\'an, Juan Garc{\'\i}a-Bellido}
\address{Departamento de F\'\i sica Te\'orica \ C-XI, 
Universidad Aut\'onoma de Madrid, 
\\ Cantoblanco, 28049 Madrid, Spain} 

\author{Julien Lesgourgues}
\address{Laboratoire d'Annecy-le-vieux de Physique Th\'eorique, \\
BP110, F-74941 Annecy-le-Vieux Cedex, France}
 
 
\begin{abstract}
The axion is one of the best motivated candidates for
  particle dark matter. We study and update the constraints imposed by
  the recent CMB and LSS experiments on the mass of axions produced by
  the misalignment mechanism, as a function of both the inflationary
  scale and the reheating temperature. Under some particular although
  not unconventional assumptions, the axionic field induces too large
  isocurvature perturbations.  Specifically, for inflation taking
  place at intermediate 
  energy scales, we derive some
  restrictive limits which can only be evaded by assuming an efficient
  reheating mechanism, with $\trh>10^{11}$ GeV. 
  Chaotic inflation with a quadratic potential is still 
  compatible with the axion scenario, provided that the Peccei-Quinn scale 
  $f_a$ is close to $10^{10}$ or $10^{11}$ GeV.
  Isocurvature bounds eliminate the possibility of
  a larger $f_a$ and a small misalignment angle. We find that
  isocurvature constraints on the axion scenario must be taken into
  account whenever the scale of inflation is above $10^{12}$ GeV;
  below this scale, axionic isocurvature modes are too small to be
  probed by current observations.
\end{abstract}

\maketitle

\section{Introduction} 

The fact that the strong sector of the Standard Model conserves the 
discrete symmetries P and CP while the electroweak sector doesn't, 
also known as the \textit{strong CP problem}, is considered a serious 
puzzle for modern particle physics~\cite{reviews}.  The most elegant 
and compelling solution to this problem was proposed in 1977 by Peccei 
and Quinn~\cite{PQ} with the introduction of a new $U(1)_{\rm{PQ}}$ 
global symmetry at high energies.  
The Peccei-Quinn (PQ) scalar field
$\psi=\rho/\sqrt{2}\,e^{i\Theta}$ has a potential of the form
\be V(\psi)= {\lambda\over4}\left(\rho^2-{f_a^2\over N^2}\right)^2 =
V_{\rm PQ} - \half m_\psi^2 \rho^2 + {\lambda\over4} \rho^4\,, \ee
where $N$ is the number of degenerate QCD vacua associated with the
color anomaly of the PQ symmetry.
Spontaneous symmetry breaking (SSB)
occurs when the energy density of the universe falls below $V_{\rm
PQ}$ and the field acquires a vacuum expectation value (VEV)
$\rho=f_a/N$. The axion~\cite{axion} is then
the Goldstone boson of the broken PQ symmetry, \be a(\vec
x)=\frac{f_a}{N}\Theta(\vec x)\,.  \ee 
Note, however, that SSB is effective only when the
typical fluctuations on $\delta\rho$ are smaller than $f_a/N$. If
either $T$ or $\hinf$ are of order $f_a/N$ at reheating or during
inflation, thermal~\cite{Harari} or quantum~\cite{LythStewart2}
fluctuations (respectively) will modify the effective potential and
restore the PQ symmetry, with a coherence length
comparable to the Hubble scale at that time.

As the universe expands, its
energy decreases to about $\Lambda_{\rm QCD} \sim 200$~MeV, 
and non-perturbative
instanton effects tilt the previously flat axion potential,
explicitly breaking the residual symmetry~\cite{witten}: 
\be\label{potential} V(a)\simeq \ma^2{f_a^2\over
N^2}(1-\cos\Theta)~. \ee The axion
field acquires a mass about the minimum of the potential that depends
on the temperature in the vicinity of $T \sim \Lambda_{\rm QCD}$
as~\cite{Gross:1980br,Fox:2004kb} 
\be \label{mass_timedep}\ma(T)\simeq m_a(0) C
\left(\frac{\Lambda_{\rm QCD}}{200\rm{MeV}}
\right)^{\frac{1}{2}}\left(\frac{\Lambda_{\rm{QCD}}}{T}\right)^4\,,
\ee where $C$ is a model-dependent factor calculated in Refs.\
\cite{Gross:1980br,Fox:2004kb} to be of the order of $C\simeq 0.018$.
The mass finally reaches its asymptotic zero-temperature value
$m_a \equiv m_a(0)$:
\begin{equation}\label{mass}
m_a = {\sqrt z\over1+z}\,{f_\pi m_\pi\over f_a} =
6.2\,\mu{\rm eV}\,\left({10^{12}\ {\rm GeV}\over f_a}\right)\,. 
\end{equation} 
Here $z=m_u/m_d\simeq0.56$ is the mass ratio of up to down quarks,
while $m_\pi$ and $f_\pi$ are respectively the pion mass and decay constant.
The field equation of motion is
\begin{equation}
\label{axion_field_eq} 
\ddot{a}   + 3H \dot{a} +V'(a)+ {1\over R^2} \nabla^2 a =0\,, 
\end{equation}  
where $V'(a)=\partial V/\partial a$, $\nabla^2$ is the comoving  
laplacian, and $R$ is the scale factor of the universe.   
If the axion field is initially 
displaced from the minimum of the potential when it acquires 
a mass, it starts oscillating in the potential described by 
Eqs.~(\ref{potential}), (\ref{mass_timedep}), (\ref{mass}).

The axion coupling to the rest of matter is inversely 
proportional to the symmetry  
breaking scale~\cite{curr,PDG}:
\begin{equation}\label{couplings} 
{\cal L}_{\rm int} = g_\gamma\,{\alpha\over\pi}\,{a(x)\over f_a}\,
\vec E\cdot\vec B +i\,g_f\,m_f\,{a(x)\over f_a}\,\bar f\,\gamma_5\,f\,, 
\end{equation} 
where $\alpha$ is the electromagnetic coupling constant, $\vec E$ and
$\vec B$ are the electric and magnetic fields, $f$ is a generic
fermion and for each species $g_i$ is a model-dependent coefficient of
order one.  The main distinction between the various axion models
comes from their coupling with electrons; for the ``hadronic'' models,
such as the KSVZ model~\cite{KSVZ}, one has $g_e=0$, while the
tree-level coupling does not vanish for the non-hadronic DFSZ
models~\cite{DFSZ}. The other couplings are of the same order.  For
example, $g_\gamma=-0.36$ in the DFSZ model, while $g_\gamma=0.97$ in
the KSVZ model.

In the currently accepted \textit{invisible axion} model, 
the scale $f_a$ is in principle 
arbitrary, well above the electroweak scale so that the axion
coupling to matter is weak enough to pass undetected, for the moment. 
There are at present several experiments searching for the axion 
in the laboratory, like ADMX~\cite{ADMX} and CAST~\cite{CAST}, which
have recently reported bounds on the axion coupling to  
matter~\cite{bounds}. 
 
Since in the large $f_a$ limit the axion remains effectively decoupled
from other species, its fluctuations during inflation could induce
isocurvature perturbations in the CMB anisotropy spectrum which
would in principle be observable
today~\cite{Axenides:1983hj,Lindeaxion,SeckelTurner,hybrid,TurnerWilczek,LindeLyth,Lyth:1991ub,Shellard:1997mf}. In
this paper, we review the consequences of up-to-date bounds on the allowed
isocurvature fraction and CDM density for the axion mass.
 
The paper is organized as follows: in Section \ref{production} we
review the different ways in which axions could be produced, with a
particular emphasis on the misalignment angle mechanism.  In Section
\ref{isocurvature} we study the induced isocurvature perturbations. 
In Section \ref{window} we present the various constraints bounding the
axionic window. 
Finally, our results are summarized and discussed in
Section \ref{discussion}.
 
\section{Production mechanisms}\label{production} 
 
Axions are produced in the early Universe by various mechanisms.  Any
combination of them could be the one responsible for the present axion
abundance. We will briefly describe here the different production
channels. For detailed reviews see Refs.~\cite{reviews,PDG}.
 
\subsection{Thermal production} 

If the coupling of axions to other species is strong enough
(i.e. $f_a$ low enough), axions may be produced thermally in the early
universe and could significantly contribute to the current dark matter
component of the universe with a relic density~\cite{KT}:
\begin{equation}\label{thermal3}
\Oa^{\rm ther} h^2=\frac{\ma}{130\,\rm{eV}}\left(\frac{10}{g_{*,F}}
\right)\,,
\end{equation}
with $\Oa\equiv\rho_a/\rho_c=8\pi G\rho_a/3H_0^2$ and $g_{*,F}$ is the
number of degrees of freedom at the temperature at which the axions
decouple from the plasma.
The current WMAP bound on the dark matter density~\cite{WMAPIII}
\be
\Ocdm h^2=0.112^{+0.003}_{-0.006}\,,
\ee together with 
Eq.~(\ref{thermal3}), imposes a bound on the axion mass $\ma<14.5$ 
eV. As we will see in Section \ref{supernovae},  
this bound is overseeded by astrophysical 
data which forbid a mass range of $0.01\,\rm{eV}< m_a <200$ keV for 
the DFSZ axion~\cite{reviews,PDG}, implying that the relic density of
thermally produced DFSZ axions is completely negligible. 

Hadronic axions are not so tigthly constrained by astrophysical data
because they do not take part (at tree level) in the processes that
cause the anomalous energy loses in stars such as $\gamma+e^-
\rightarrow e^- +a$ or the Primakoff effect \cite{reviews}.
However, using the fact
that thermally produced axions behave as warm dark matter, the authors
of Ref.~\cite{Hannestad:2005df} derived a model-independent bound
$m_a<1.05$~eV showing that in any case $\Omega_{\rm a}^{\rm ther} \ll
\Omega_{\rm cdm}$. We will therefore ignore from now on the thermal
axion contribution to dark matter, and work in the limit in which
axions are completely decoupled from the rest of matter.
 
\subsection{Production via cosmic strings} 

We already mentioned that the PQ symmetry could be restored at high
energy. After each symmetry restoration phase, a SSB can produce a
population of axionic cosmic strings at the following epochs:

\begin{itemize}
  \item {\it after inflation:} if the scale $f_a/N$ is below the
    reheating temperature of the universe, the PQ symmetry is restored
    at reheating by thermal fluctuations. Axionic cosmic strings
    are produced later, when the temperature drops below
    $f_a/N$~\cite{Harari}.  These strings typically decay into axion
    particles before dominating the energy density of the
    universe. The axions produced this way are relativistic until the
    QCD transition, where they acquire a mass and become
    non-relativistic. Eventually these axions may come to dominate the
    energy density after equality, in the form of cold dark
    matter. Estimates of their present energy density vary depending
    on the fraction of axions radiated by long strings versus string
    loops. Three groups have studied this issue and found agreement
    within an order of magnitude~\cite{Shellard,Harari,Yamaguchi},
\begin{equation} 
\Oa^{\rm str} h^2 \simeq 4\Delta_{\rm QCD}\, 
\left({1\mu{\rm eV}\over\ma}\right)^{1.18}\,,
\label{axion_density_from_strings}
\end{equation} 
where $\Delta_{\rm QCD}=3^{\pm1}$ is a ``fudge factor'' which takes
into account all the uncertainties in the QCD phase transition.
Similar bounds were found in \cite{Khlopov:1999tm} following a very
different approach.

\item {\it at the end of inflation:} the PQ symmetry is restored during
  inflation whenever the typical amplitude of quantum fluctuations
  $\hinf/2\pi$ exceeds the symmetry breaking scale $f_a/N$~\cite{LythStewart2}.
  If the inequality
\begin{equation}
\frac{H_{\rm inf}}{2 \pi} > \fN
\label{sym_res} 
\end{equation}
holds throughout inflation, comic strings will be produced at the very
end of this stage. The mechanism by which cosmic strings are produced
at the end of inflation is very different from that of a thermal phase
transition and could affect the number of infinite strings, and thus
the approach to the scaling limit and the relic density of axions.
A detailed analysis of the axionic string production after inflation
and the corresponding estimate of the relic density of axions lies somewhat
out of the scope of this paper and is left for future work.
In the next sections
and in Figs.~\ref{fig:bounds}, \ref{fig:bounds2}, we will tentatively
assume that the scaling limit is approached and that the corresponding
relic density of axion is still given by
Eq.~(\ref{axion_density_from_strings}), but the actual density could
be significantly smaller. In any case, we will see that this bound is
overseeded by the one derived from isocurvature modes.

\item {\it during inflation:} the PQ SSB could occur during inflation,
  when $H_{\rm inf}/2\pi$ falls below $f_a/N$. In this case, one expects
  axionic strings to be diluted during the remaining inflationary
  stage, and the relic density will be suppressed by an additional
  factor, $\exp(N_{\rm SSB})$, where $N_{\rm SSB}$ is the number of
  e-folds between PQ symmetry breaking and the end of inflation. As a
  consequence, this density should be negligible today, unless $N_{\rm
    SSB}$ is fine-tuned to small values by assuming that $f_a/N$ is
  very close to the Hubble rate at the end of inflation.

\item {\it before inflation:} if, for instance, the PQ symmetry is
  restored at very high energy and breaks before inflation is turned
  on e.g. at low scales, the axionic strings produced in that way,
  as well as the possible axions into which they may have decayed,
  will be diluted by inflation and can be safely neglected.

\end{itemize}

In addition, if $N>1$, axionic domain walls could be generated
during the explicit symmetry breaking occurring around the QCD scale.
Their energy density would end up as the dominant component of the
universe~\cite{wall} unless one of the possible vacua turned out to be
slightly non-degenerate.  Since this would require a considerable
amount of fine tunning, we will assume in the rest of this paper that
$N=1$. A detailed discussion of this issue is given in
Ref.~\cite{chs}.
 
\subsection{Generation via misalignment angle}
\label{sec:mis}
 
When the PQ symmetry is explicitly broken by instanton effects, the
phase of the field $\Theta$ may not be at the minimum of its
potential.  As explained above, $a$ (or $\Theta$) is a massless field
during inflation and thus it fluctuates quantum-mechanically.  If the
typical amplitude of quantum fluctuations is large enough, $\Theta$
could take different values in different points of our observable
universe after inflation, with a flat probability distribution in the
range $[-\pi,\pi]$; otherwise, it could remain nearly homogeneous.  In
both cases, at the time of the QCD transition, the (local or global)
value of the misalignment angle $\Theta_{\rm QCD}$ can differ from zero,
leading to the sudden appearance of a potential energy
$m_a^2 f_a^2 (1-\cos \Theta_{\rm QCD})$.
After the explicit symmetry breaking, the axion energy reads 
\begin{eqnarray}\nonumber
\rho_{\rm{a}} &=& \half\dot{a}^2+ 
\frac{(\vec{\nabla} a)^2}{2R^2} + 
\ma^2f_a^2(1-\cos\Theta) \\ \label{axion_rho_aesb} 
&\simeq& \half f_a^2\Big(\dot{\Theta}^2+
\frac{(\vec{\nabla} \Theta)^2}{R^2} + 
\ma^2 \Theta^2\Big) 
\hspace{2mm}\textrm{for small $\Theta$}.  
\end{eqnarray}
Actually, the gradient
energy can be safely neglected in Eq.\ (\ref{axion_rho_aesb}). Indeed,
even in the case in which the axion is maximally inhomogeneous, 
i.e. when the phase $\Theta$ is equally distributed in the range 
$[-\pi,\pi]$ in our observable universe at the end of inflation, it is 
straightforward to show that at any later time the coherence length 
(the physical size of the ``homogeneity patches'' for $a$) is always 
of the order of the Hubble radius. This can be checked e.g. by solving 
the equation of motion (\ref{axion_field_eq}) in Fourier space. As a 
consequence, and recalling that $a$ is defined in the range $[-\pi 
f_a, \pi f_a]$, the typical size of the gradient $(\vec{\nabla} a / R)$ is 
given by $f_a H$. So, the gradient energy scales as $H^2 \propto 
R^{-4}$ during radiation domination, and at the time of the QCD 
transition it is at most of the order of $(f_a H_{\rm QCD})^2$.  A 
quick estimate gives $H_{\rm QCD} \sim 10^{-11}$eV, while in the 
following we will always consider values of the axion mass much larger 
than this. So, when the axion mass is ``switched on'', the gradient 
energy is negligible with respect to the potential energy $V \sim (f_a 
m_a)^2$. 
 
When $m_a(T)$ grows suddenly to values much bigger than $H_{\rm QCD}$,
the field quickly rolls down towards the mini\-mum of the potential
and starts oscillating with an energy density given in first
approximation by \be\label{rho} \rho_a= \left\langle
\frac{1}{2}f_a^2\Big(\dot{\Theta}^2 + m_a^2 \Theta^2\Big)
\right\rangle = \frac{1}{2} f_a^2 m_a^2
\left\langle\Theta_{\rm QCD}^2\right\rangle \left( \frac{R_{\rm QCD}}{R}
\right)^3~. \ee However, for a precise estimate, it is necessary to
take into account the time dependence of the mass. This was done in
Ref.~\cite{Turner1986}, which found an extra factor $f_c \simeq 1.44$
in Eq.~(\ref{rho}).  Finally, using the conservation of the number
density $\na=\rho_a/\ma(T)$ in a comoving volume, one finds a relic
axion density \be\label{axion_relic_mis} \Oa h^2\simeq 7.24
\,g_{*,1}^{-5/12}\langle\Theta_{\rm QCD}^2\rangle
\left(\frac{200\rm{MeV}}
{\Lambda_{\rm{QCD}}}\right)^{\frac{3}{4}}\left(\frac{1\mu\rm{eV}}
{m_a}\right)^{\frac{7}{6}}\,, \ee where $g_{*,1} \simeq 61.75$ is the
number of relativistic degrees of freedom in the universe at the
temperature when the axion starts oscillating. Note that in
Eq.~(\ref{axion_relic_mis}), the spatial average $\langle\Theta_{\rm
QCD}^2\rangle$ of the initial misalignment squared angle over the
observable universe is not given by any field theoretical reasoning,
but by considerations on the stochastic behavior of the field during
inflation.  Using the Fokker-Planck equation, it is easy to show that
~\cite{LindeLyth,Lyth:1991ub,LythStewart2} \be\label{delta2} \langle
(\Theta_{\rm in}-\Theta_{\rm QCD})^2\rangle^{1/2} \sim {\hinf\over2\pi
f_a}\times\sqrt{\Delta N_{\rm QCD}}\,, \ee where $\Theta_{\rm in}$ is
the average value of the phase in our inflationary patch at the time
of Hubble crossing for the observable universe ($N_{\rm
obs}$ e-folds before the end of inflation), and $\Delta N_{\rm QCD}
\simeq 30$ is the number of e-folds between $N_{\rm obs}$ and the time
of Hubble exit for the comoving scale which re-enters the horizon when
$H=H_{\rm QCD}$.

At this point, we see that two situations can occur. First, if $f_a
\gg \hinf$, the right-hand-side in equation (\ref{delta2}) can be much
smaller than one at the end of inflation; then, the axion field is
essentially homogeneous, and the background value 
$\Theta_{\rm QCD}=\Theta_{\rm in}$ in our
universe is random but unique.  Second, if $f_a \leq \hinf$, the
right-hand-side can be of order one or larger, which means that the
Brownian diffusion of the axion is complete, and the misalignment
angle at the QCD scale is randomly distributed with a flat probability
distribution in the range $[-\pi,\pi]$. Note that in this case, the
quantum perturbations of the radial part of the PQ field are also
large during inflation. In both
cases, the mean energy density of the axion around $\Lambda_{\rm QCD}$
is proportional to
\be\label{meantheta}
\langle\Theta_{\rm QCD}^2\rangle=\frac{1}{2\pi}\int_{-\pi}^{\pi}\alpha^2
d\alpha=\frac{\pi^2}{3}\,, \ee where the average should be understood
as holding over many realizations of the universe in the case of a
nearly homogeneous $\Theta_{\rm QCD}$, or over our present Hubble radius in
the case of complete diffusion.
 
Up until this point we have ignored the anharmonic corrections that 
could arise from the possibly large value of $\Theta_{\rm QCD}$. The 
calculations have been made using the approximation $(1-\cos\Theta) 
\simeq\frac{1}{2}\Theta^2$, which is obviously not valid for large 
angles. Including anharmonic corrections, one finds an enhancement factor
1.2 in Eq.~(\ref{meantheta}).
 
\section{Isocurvature perturbations from axion fluctuations} 
\label{isocurvature} 
 
If the PQ symmetry is spontaneously broken during inflation, while the
scale of inflation is much higher than that of the quark-hadron
transition, the flat direction associated with the massless
Nambu-Goldstone boson will be sensitive to de Sitter quantum
fluctuations. Indeed, quantum fluctuations are imprinted into every
massless scalar field present during inflation, with a nearly scale
invariant spectrum, \be \label{de_sitter_fluct} \langle |\delta
a(k)|^2 \rangle
=\left(\frac{\hinf}{2\pi}\right)^2\frac{1}{k^3/2\pi^2}\,.  \ee If the
scale of inflation is high enough, $\hinf/2\pi > f_a$, it is possible
that quantum fluctuations of the radial part of the PQ field restores
the symmetry~\cite{LythStewart2}. This symmetry restoration could have
very different implications for cosmological perturbations than a
possible thermal symmetry restoration taking place after
inflation. Indeed, the effective mass-squared $V''(\rho)$ at the false
vacuum $\rho=0$ is much smaller than the Hubble rate, $V''(\rho) =
\hinf^2/48\pi^2 \ll \hinf^2$. So, the PQ field behaves like a light
complex field during inflation. The symmetry is restored in the sense
that an average over a scale much larger than the coherence length
$\lambda_{\psi}(t)$ of the field would give $\langle \psi \rangle=0$.
However, $\lambda_{\psi}(t)$ is of the same order as the Hubble radius
$c/H(t)$ at a given time. In comoving space, the coherence length
decreases by a huge factor $e^{\Delta N}$ during inflation, and the
evolution of the field can be seen as a stochastic process of
fragmentation into smaller and smaller homogeneity patches.  But at
the time when our observable universe crosses the Hubble scale, the PQ
field is still nearly homogeneous inside our patch. Its quantum
fluctuations become frozen beyond the horizon, and could thus leave a
long wave perturbation which would still be described by
Eq.~(\ref{de_sitter_fluct}). A detailed proof of this highly
non-linear process requires a lattice simulation, whose analysis we
leave for a future publication. In this paper, we will conservatively
assume that isocurvature perturbations are erased when $\hinf/2\pi >
f_a$, as has been assumed so far in the rest of the
literature~\cite{LythStewart2}.

The axion field perturbations $\delta a$ do not perturb the total
energy density, first because the potential energy is exactly zero,
and second because, as explained above, the gradient energy of the
axion cannot exceed $\sim (f_a H_{\rm inf})^2$; for $f_a \ll M_P$ this
is much smaller than the total energy density $(3/8 \pi) M_P^2 H_{\rm
inf}^2$.  Since the total energy density is unperturbed by these
fluctuations during inflation, they are of \textit{isocurvature} type,
and manifest themselves as fluctuations in the number density of
axions~\cite{Axenides:1983hj,Lindeaxion,hybrid,TurnerWilczek,SeckelTurner},
\be \delta\left(\frac{\na}{s}\right)\neq 0\,.  \ee In the absence of
thermal symmetry restoration after inflation, i.e. if the temperature
of the plasma does not reach $f_a$, the axion does not couple
significantly to ordinary matter. Its perturbations remain truly
isocurvature~\cite{Weinberg} and may contribute as such to the
temperature anisotropies of the
CMB~\cite{Polarski:1994rz,Garcia-Bellido:1995qq,Gordon:2000hv}.  This
isocurvature mode is expected to be completely uncorrelated to the
usual adiabatic mode seeded by the quantum fluctuations of the inflaton.
 
Let us assume that the Universe contains photons ($\gamma$),
approximately massless neutrinos ($\nu$), baryons (b), axions (a),
ordinary CDM such as neutralinos (x) and a cosmological constant. In
the following, the subscript cdm will denote the {\em total} cold dark
matter component, so that $\Omega_{\rm cdm} = \Omega_{\rm a} +
\Omega_{\rm x}$.  For the isocurvature mode, the perturbation
evolution starts from the initial condition $\frac{3}{4}
\delta_{\gamma}=\frac{3}{4} \delta_{\nu}=\delta_{\rm b}=\delta_{\rm x}
\simeq 0$ and $\delta_{\rm a}={\cal S}_{\rm a}$, where ${\cal S}_{\rm
a}$ is the gauge invariant entropy perturbation \be\label{entropy_def}
\mathcal{S}_{\rm a}=\frac{\delta(n_{\rm a}/s)}{(n_{\rm
a}/s)}=\frac{\delta n_{\rm a}}{n_{\rm a}}-3\frac{\delta T}{T}\, \ee
(indeed, after the QCD transition, the axion fluid is non-relativistic
with $\rho_{\rm a} = m_{\rm a} n_{\rm a}$, so $\delta_{\rm a}=(\delta
n_{\rm a})/n_{\rm a}$; furthermore, the fact that $\delta_{\rm a} \gg
\delta_{\gamma}$ implies $\delta n_{\rm a}/n_{\rm a} \gg 4 \delta T / 
T$ and ${\cal S}_{\rm a} = \delta n_{\rm a}/n_{\rm a}=\delta_a$).  
It is equivalent to consider the perturbations of a
single cold dark matter fluid, obeying now to the initial condition
$\delta_{\rm cdm}=R_{\rm a} \delta_{\rm a} + (1-R_{\rm a})
\delta_x=R_{\rm a} {\cal S}_{\rm a}$. If we compare with the initial
condition for a usual ``Cold Dark matter Isocurvature'' (CDI) model,
given by $\delta_{\rm cdm}={\cal S}_{\rm cdm}$, we see that the
axionic isocurvature solution is equivalent to the CDI solution with
${\cal S}_{\rm cdm} = R_{\rm a} {\cal S}_{\rm a}$.  In other words, an
axionic model with axionic fraction $R_{\rm a}=\Omega_{\rm
a}/\Omega_{\rm cdm}$, initial curvature spectrum $\langle {\cal R}^2
\rangle$ and initial entropy spectrum $\langle {\cal S}_{\rm a}^2
\rangle$ is strictly equivalent to a mixed adiabatic+CDI model with
the same curvature spectrum and $\langle {\cal S}_{\rm cdm}^2 \rangle
= R_{\rm a}^2 \langle {\cal S}_{\rm a}^2 \rangle$.
 
Let us now relate the curvature and entropy power spectrum to the quantum 
fluctuations of the inflaton and axion field during inflation. 
For the adiabatic mode, it is well-known that the curvature power 
spectrum reads  
\begin{equation} 
\langle |{\cal R}(k)|^2 \rangle = \frac{2 \pi H_k^2}{k^3 M_P^2 \epsilon_k} 
\label{curvature_amp} 
\end{equation} 
where $\epsilon$ is the first inflationary slow-roll parameter 
\cite{LLbook} and the subscript $k$ indicates that quantities are 
evaluated during inflation, when $k=aH$.  In first approximation this 
spectrum is a power-law with a tilt $n_{\rm ad}$ depending also on the 
second slow-roll parameter $\eta$ \cite{LLbook}, 
\begin{equation} 
n_{\rm ad}=1 -6 \epsilon_k + 2 \eta_k\,. 
\end{equation} 
For the isocurvature mode, using the axion perturbation 
spectrum of Eq.\ (\ref{de_sitter_fluct}), we obtain 
\begin{equation} 
\langle |{\cal S}_{\rm a}(k)|^2 \rangle = \left\langle \left|  
\frac{\delta n_{\rm a}}{n_{\rm a}} \right|^2 \right\rangle 
= 4 \left\langle \left| \frac{\delta a}{a} \right|^2 \right\rangle 
= \frac{2 H_k^2}{k^3 f_a^2 \langle \Theta_{\rm QCD}^2 \rangle}\,. 
\end{equation} 
This power spectrum can be approximated by a power-law with a tilt 
\begin{equation} 
\niso=1 -2 \epsilon_k\,,
\end{equation} 
which is related to the tilt $n_t$ of tensor perturbations,
$\niso = 1 + n_t$.

The relative amplitude of isocurvature perturbations at a given pivot scale
in adiabatic+CDI models is often parametrized
as~\cite{Crotty:2003rz,Beltran:2004uv,Beltran:2005xd,Beltran:2005gr} 
\begin{equation} 
\alpha = \frac{\langle |{\cal S}_{\rm cdm}(k)|^2 \rangle} 
{\langle |{\cal S}_{\rm cdm}(k)|^2 \rangle + 
\langle |{\cal R}(k)|^2 \rangle}~. 
\end{equation} 
Since the axionic model is equivalent to an adiabatic+CDI model, we 
can still use the same parametrization. The parameter $\alpha$ is 
related to fundamental parameters by 
\begin{equation} 
\label{alpha} 
\alpha = \frac{R_{\rm a}^2 \langle |{\cal S}_{\rm a}(k)|^2 \rangle}{R_{\rm a}^2 \langle
|{\cal S}_{\rm a}(k)|^2 \rangle + 
\langle |{\cal R}(k)|^2 \rangle} \simeq  
\frac{R_{\rm a}^2 M_P^2 \epsilon_k}{\pi f_{\rm a}^2 \langle \Theta_{\rm QCD}^2 \rangle} 
\end{equation} 
where in the last equality we assumed $\alpha \ll 1$.

\section{The axionic window} 
\label{window} 
 
We now describe the generic constraints on $\fa$ coming from different 
cosmological and astrophysical considerations. 
Some of them are generic and apply to every production mechanism or 
inflationary scenario while some others are rather model dependent. 
More precisely, we will enumerate the different bounds on the axion 
parameter space $(M_{\rm inf},\,f_a)$ associated with the
misalignment mechanism of axion production during inflation, 
$M_{\rm inf}$ being the energy scale during inflation 
\be\nonumber 
M_{\rm inf}\equiv\sqrt{\frac{\MP H_{\rm inf}}{\sqrt{8\pi/3}}}\,,
\ee 
and $M_{\rm P}=G^{-1/2}$ is the Planck mass.  The
explicit implications of these bounds are presented on 
Figs.~\ref{fig:bounds}, \ref{fig:bounds2}.
 
\subsection{The scale of inflation} 
\label{scale}
The non-detection of tensor modes imposes a 
model-independent constraint on 
the inflationary scale,
\begin{equation} 
\label{tensorlimit} 
M_{\rm inf} < 3 \times10^{16}\ {\rm GeV}\,, 
\end{equation}
which follows from the current limit on the tensor to scalar ratio 
\cite{WMAPIII}
\begin{equation} 
r \sim \left(4 \times10^4\,H_{\rm inf}\over M_{\rm P}\right)^2 <  
0.3 \hspace{3mm} {\rm at}\ 95\%\ {\rm c.l.}\,. 
\end{equation} 
In Figs.~\ref{fig:bounds}, \ref{fig:bounds2}, 
this constraint corresponds to the upper 
hatched forbidden region.

\subsection{Supernova 1987A bounds}  
\label{supernovae} 
 
The observed neutrino luminosity from supernova 1987A 
imposes a bound on the axion luminosity that is saturated for 
$10^{-2}\,{\rm eV}<\ma < 2\,{\rm eV}$ \cite{PDG}. 
Other astrophysical and laboratory searches rule out an axion heavier 
than 1 eV (see \cite{reviews,PDG} for a detailed discussion).  
Therefore, we have: 
\begin{equation} 
m_a < 10^{-2} - 10^{-3} \ {\rm eV}\,, 
\end{equation} 
or equivalently, $f_a > 10^9 - 10^{10}$ GeV.  The forbidden region is
(blue-)shaded in Figs.~\ref{fig:bounds}, \ref{fig:bounds2}.
 
\subsection{Axionic cosmic strings production} 
\label{strings}

As mentioned before, in the case of symmetry restoration at high
energy, axions can be produced from the decay of cosmic strings,
and we should impose a bound on their relic density
\begin{equation} 
\omega^{\rm str}_{\rm a} \leq \omega_{\rm cdm}~,
\label{cosmic_strings}
\end{equation}
where $\omega_i$ stands for $\Omega_i h^2$ and 
$\Oa^{\rm str}$ is taken from
Eq.~(\ref{axion_density_from_strings}). 
Taking $\Omega_{\rm cdm}\,h^2 < 0.123$ (see Eq.(\ref{ocdm_bound})) we get
\begin{equation}
f_a < 1.25 \times 10^{11} {\rm GeV} ~.
\end{equation}
This inequality must be
imposed in two cases: $f_a > (H_{\rm inf}/2\pi)$, corresponding to
symmetry restoration during inflation\footnote{Note that for
  simplicity, we impose this condition as if $\hinf$ was constant at
  least during the observable e-folds of inflation (typically, the
  last sixty e-folds). In principle, the amplitude of quantum
  fluctuations $(\hinf/2 \pi)$ could fall below $f_a$ precisely during
  the observable e-folds, see e.g. \cite{Lindeaxion}, but we will
  ignore this possibility here.}~\cite{LythStewart2}, and $f_a > T_{\rm rh}$,
corresponding to symmetry restoration after reheating~\cite{Harari}. Actually,
 it is worth mentioning that
  thermal corrections induce a positive mass-squared term $m_{\rm
    eff}^2 = T^2_{\rm rh}/12$, so the precise condition for symmetry
  restoration is 
\be
\label{thermal_resto}
T^2_{\rm rh}/12 \gg \lambda f_a^2~.
\ee 
Therefore, the usual statement
  that symmetry is restored whenever $f_a > T_{\rm rh}$ assumes that 
  the self-coupling constant $\lambda$ is of the order of 0.1. We will proceed with this assumption, but one should keep in mind that the exact condition is model-dependent.

The actual reheating temperature $T_{\rm rh}$
is still unknown. This is the temperature at which the inflaton 
decays, once its half life has been exceeded by the age of the 
universe,  that is, when $H\sim\Gamma$. 
Most of the thermal energy comes from perturbative decays of the 
inflaton and, assuming that the decay products are strongly 
interacting at high energies, we can estimate the reheating 
temperature as 
\begin{equation} 
T_{\rm rh} \simeq 0.1\,\sqrt{\Gamma M_P} \simeq 0.02\,h_{\rm eff}\, 
\sqrt{m\, M_P}  \leq 2\times10^{11}\ {\rm GeV}\,, 
\end{equation} 
where $\Gamma=h^2_{\rm eff}m/8\pi$ is the inflaton decay rate, which 
is typically proportional to the inflaton mass, $m$, with $h_{\rm 
eff}\leq 10^{-3}$, in order to prevent radiative corrections from 
spoiling the required flatness of the inflaton 
potential~\cite{Lindebook}.  This estimate shows the generic 
inefficiency of reheating after inflation, where the scale of 
inflation could be of order $10^{15}$ GeV and the reheating 
temperature ends being many orders of magnitude lower. For instance,
for Starobinsky type inflation, the weak gravitational couplings
give a reheating temperature of order $\trh \sim 10^9$ GeV, while
in chaotic inflation models, typical values are of order 
$10^{10} - 10^{11}$ GeV. On the other hand, in certain low scale
inflationary models, such as hybrid inflation, the efficiency of
reheating can be significant because the rate
of expansion at the end of inflation is much smaller than any other
mass scale and the inflaton decays before the universe has time to
expand, therefore all the inflaton energy density gets converted into
radiation.

We can
parametrize the effect on the rate of expansion by introducing an
efficiency parameter, $\epsilon_{\rm eff}$, such that $H_{\rm rh} =
\epsilon_{\rm eff} \,H_{\rm end}$. Values of $\epsilon$ range from
$10^{-13}$ for Starobinsky inflation, to order one for very low
scale inflation. If the reheating temperature is higher than $f_a$
it could eventually lead to a restoration of the PQ symmetry. The
subsequent spontaneous symmetry breaking would generate axionic cosmic
strings that would not be diluted away by inflation.

In summary, if the PQ symmetry is restored by thermal fluctuations
after reheating, i.e.
\begin{equation}\label{reheating}
T_{\rm rh} = 0.1 \sqrt{H_{\rm rh}M_{\rm P}} =
0.1 \sqrt{\epsilon_{\rm eff} H_{\rm inf}M_{\rm P}} > f_a\,,
\end{equation}
then we must impose the condition (\ref{cosmic_strings}).  In
Figs.~\ref{fig:bounds}, \ref{fig:bounds2}, we have distinguished two
cases: one in which the process of reheating the universe is very
inefficient ($\epsilon_{\rm eff} \leq 10^{-12}$), and there is no
thermal restoration of the PQ symmetry after inflation, and another
one in which $\epsilon_{\rm eff}=10^{-4}$ so that the symmetry might
be restored. In both cases the constraint coming from string
production and decay corresponds to the triangular (red-)shaded
exclusion region.
 
\subsection{Cold Dark Matter produced by misalignment}\label{cdm}

As for axions produced by string decay,
the relic density of axions produced by misalignment
should not exceed the total
cold dark matter density.
From Eq.~(\ref{axion_relic_mis}) we have 
\begin{eqnarray}\label{relic_bounds}\nonumber 
\omega_{\rm a} \simeq 1.3  
\langle\Theta_{\rm QCD}^2\rangle 
\left({1\,\mu {\rm eV}\over m_a}\right)^{7/6} &=&\\
2.8 \times 10^7 \langle\Theta_{\rm QCD}^2\rangle  
\left( \frac{f_a}{M_{\rm P}} \right)^{7/6} 
&\lsim & \ \omega_{\rm cdm} \,,  
\end{eqnarray} 
where we neglected the anharmonic correction factor.
This inequality provides a stringent upper limit on $f_a$
if $\langle\Theta_{\rm QCD}^2 \rangle$ is of order one.
In particular, in the case of complete quantum diffusion during inflation,
we have seen at the end of Sec.~\ref{sec:mis}
that $\langle\Theta_{\rm QCD}^2 \rangle$ can be replaced by
$1.2 \, \pi^2/3$
and
\begin{equation}
f_a \leq 2.5\times10^{11} {\rm GeV}
\left(\frac{\omega_{\rm cdm}}{0.12}\right)^{6/7}
\label{density_constraint1}.
\end{equation}
This constraint is shown in Figs.~\ref{fig:bounds}, \ref{fig:bounds2}
as a dotted line (assuming $\omega_{\rm cdm} = 0.12$), and excludes
the light green region.

In the absence of efficient quantum diffusion, $\Theta_{\rm QCD}$
could take any nearly homogeneous value in our Universe: so, it is
possible in principle to assume that $\langle \Theta_{\rm QCD}^2
\rangle$ is extremely small (this coincidence can be justified by
anthropic considerations\footnote{There has been plausible
speculations that our presence in the universe may not be uncorrelated
with the values of the fundamental parameters in our theories. Such
anthropic arguments normally arise in terms of conditional probability
distributions of particular observables. In particular, the axion
abundance is a natural parameter that could be bounded by those
arguments, see e.g.  Refs.~\cite{Lindeaxion,Tegmark} where it is
suggested that the initial misalignment angle should be such that the
main CDM component be axionic. In this case, one has a concrete
prediction for $R_{\rm a} = \Oa/\Omega_{\rm cdm} = 1$, and therefore
the initial misalignment angle is directly related to the axion mass,
see Eq.~(\ref{relic_bounds}), \be\label{matheta} \ma = 9\,\mu {\rm
eV}\,\langle\Theta_{\rm QCD}^2\rangle^{6/7}\,.  \ee Having full
diffusion, $\langle\Theta_{\rm QCD}^2\rangle \sim 1.2\pi^2/3$, implies
$\ma \sim 30\,\mu$eV, just within reach of present axion dark matter
experiments.  On the other hand, we might happen to live in an unusual
region of the universe with an extremely low value of
$\langle\Theta_{\rm QCD}^2\rangle$, a large value of $f_a$ and still
$R_a=1$.}), and to relax the bound on $f_a$. However, the mean square
cannot be fine-tuned to be smaller than the amplitude of quantum
fluctuations at the end of inflation. Using Eq.~(\ref{delta2}), we see
that $\langle \Theta_{\rm QCD}^2 \rangle$ can only vary within the
range
\begin{equation}
30 \left( \frac{H_{\rm inf}}{2 \pi f_a}\right)^2 < \langle \Theta_{\rm QCD}^2 \rangle
< \frac{\pi^2}{3}~.
\end{equation}  
This gives a model-independent constraint
\begin{equation}
M_{\rm inf} \leq 2.5 \times 10^{15} {\rm GeV}
\left(\frac{\omega_{\rm cdm}}{0.12}\frac{30}{\Delta N}\right)^{\frac{1}{4}}
\left(\frac{f_a}{10^{12} {\rm GeV}}\right)^{\frac{5}{24}}
\end{equation}
which holds only in the region where 
\begin{equation}
30 \left( \frac{H_{\rm inf}}{2 \pi f_a}\right)^2
< \frac{\pi^2}{3}~,
\end{equation}
otherwise it should be replaced by (\ref{density_constraint1}). This bound
excludes the dark green region in Figs.~\ref{fig:bounds}, \ref{fig:bounds2}
(assuming $\omega_{\rm cdm}=0.12$ from WMAP 3rd year results).

\subsection{Isocurvature modes}\label{iso}

As mentioned before, the axionic field induces an isocurvature
component in the CMB anisotropies that must be considered when
constraining the model. In this work, we assume that axions are the
only source of isocurvature modes.  Taking expression (\ref{alpha})
for the isocurvature fraction $\alpha$, replacing $R_{\rm a}$ by
$\omega_{\rm a}/\omega_{\rm cdm}$ and using Eq.\ (\ref{relic_bounds}),
we obtain
\begin{equation} 
\alpha = \frac{0.9 \times 10^7 \epsilon_k}{\ocdm^2}  
\left( \frac{M_{\rm P}}{f_a} \right)^{5/6}~. 
\label{alpha2}
\end{equation} 
We will now fit this model to the cosmological data in order to
provide updated bounds on $\alpha$ and $\omega_{\rm cdm}$. The
analysis will also give constraints on the curvature power spectrum
$\langle |{\cal R}(k)|^2 \rangle$, from which one can derive a
relation between $\epsilon_k$ and $M_{\rm inf}$, using equation
(\ref{curvature_amp}).  Therefore, our bounds on $\alpha$ and
$\omega_{\rm cdm}$ will provide
constraints in the $(M_{\rm inf},\,f_a)$ plane.
 
We assume a flat $\Lambda$CDM universe, with 3 species of massless
neutrinos and seven free parameters, with the same notation as in
Ref.~\cite{Beltran:2004uv}: $\omega_b$, the physical baryon density,
$\ocdm$, the total cold dark matter density (an arbitrary fraction of
which is made of axions), $\theta$, the angular diameter of the sound
horizon at decoupling, $\tau$, the optical depth to reionization,
$\nad$, the adiabatic spectral tilt, $A_s=\ln[10^{10} k^3 \langle
|{\cal R}|^2 \rangle]$, the overall normalization of super-Hubble
curvature perturbations during radiation domination at the pivot scale
$k=0.002$~Mpc$^{-1}$ and $\alpha$, the isocurvature contribution (we
sampled from $|\alpha|$ instead to avoid boundary effects near the
maximum likelihood region).  Our data consists of CMB data (from WMAP
(TT, TE and EE)~\cite{WMAPIII}, VSA~\cite{VSA}, CBI~\cite{CBI} and
ACBAR~\cite{ACBAR}); large scale structure data (2dFGRS~\cite{2dFGRS}
and the SDSS~\cite{SDSS}); and supernovae data from
Ref.~\cite{Perlmutter:1998np}.  
Note that previous studies
\cite{Beltran:2004uv} indicated a very weak sensitivity of this data
to $n_{\rm iso}$, while in the present case $n_{\rm
iso}=1-2\epsilon_k$ is very close to one, since 
the data favour $\epsilon_k \ll 1$. Thus, we safely fix $n_{\rm
iso}$ to exactly one without modifying the results. 
Note also that
the isocurvature mode arises exclusively from perturbations in the
axion field while the adiabatic mode emerges from quantum fluctuations
of the inflaton. Therefore, both contributions are completely
uncorrelated and we do not need to include an extra parameter 
that would measure this contribution (such as 
$\beta$ under some parametrizations \cite{Crotty:2003rz}) in our analysis.
 
A top-hat prior probability
distribution was assigned to each parameter inside the ranges
described in table \ref{param_space}.
 
\begin{table}[t] 
\begin{center}
\begin{tabular}{|c|c|c|}
\hline
Parameter& Prior probability range & Mean ($\pm 2 \sigma$)\\
\hline
$\oB$      & (0.016,0.030) &   $0.023 \pm 0.001$ \\
$\ocdm$     & (0.08,0.16)  &   $0.123 \pm 0.008$ \\
$\theta$   & (1.0,1.1) &       $1.04 \pm 0.01$ \\
$\tau$   & (0.01,0.2) &    $0.09 \pm 0.04$ \\
$\nad$      & (0.85,1.1) &    $0.97 \pm 0.03$ \\
$A_s$       &(2.7,4.5) &    $3.1 \pm 0.1$ \\
$|\alpha|$    & ($-1$,1) &  $< 0.08$\ (95\% c.l.)  \\
\hline
\end{tabular}
\end{center}\caption{Prior probability ranges and means of the sampled parameters.}\label{param_space}
\end{table}

We used the Metropolis-Hastings algorithm
implemented by the publicly available code {\tt CosmoMC}
\cite{cosmomc} to obtain 32 Monte Carlo Markov chains, getting a total
of $1.1\times 10^5$ samples.
We find a $\chi^2 / \rm{d.o.f.}=1.01$
and the worst variance of chain means over the mean of
chain variances value is 1.04~\cite{rubin}.

\begin{figure*} 
\begin{center} 
\begin{tabular}{cccc}  
\hspace{-1cm} 
\epsfig{figure=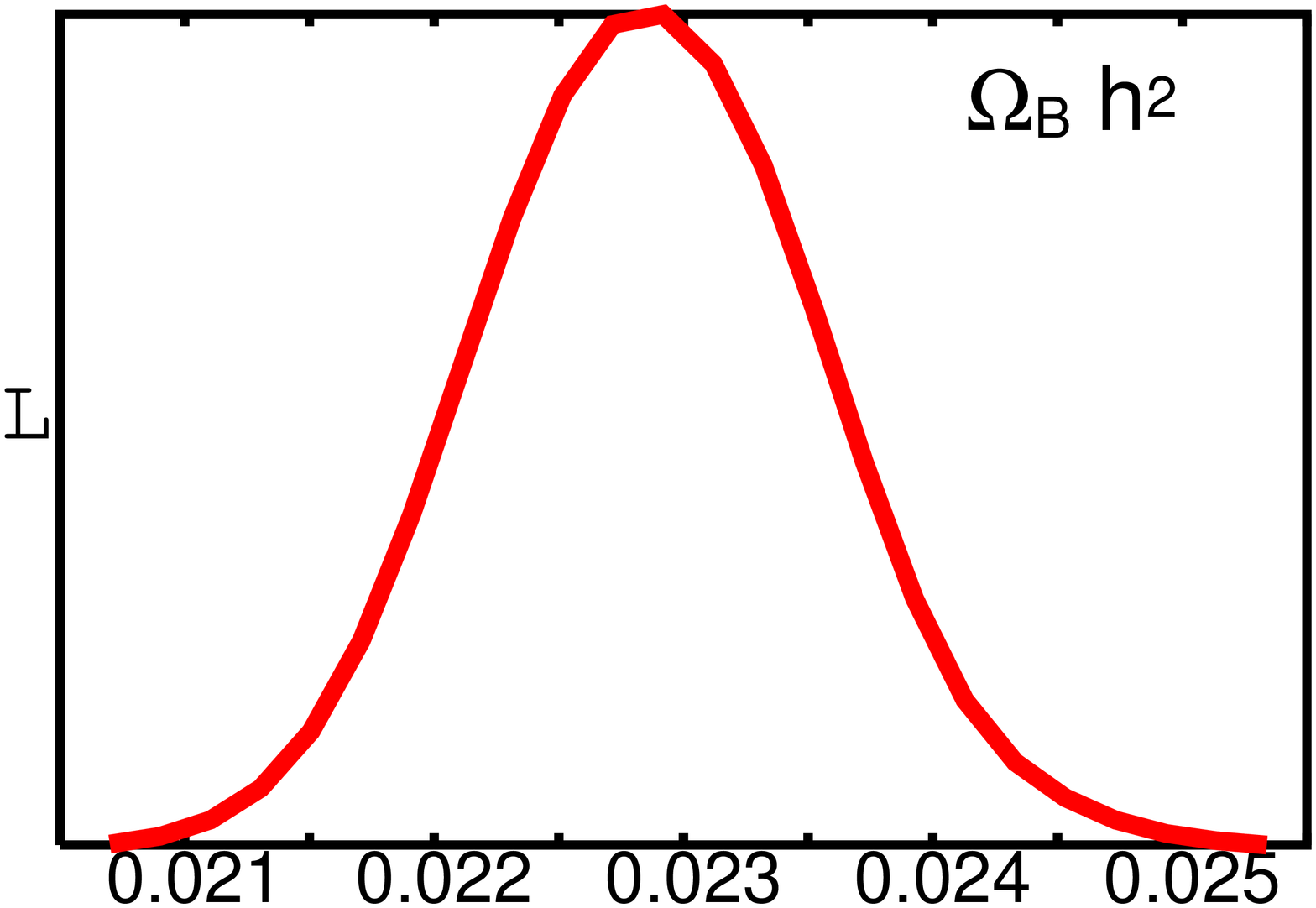,  height=3.6cm, width=3cm, angle=-90 }   & 
\hspace{.5cm} 
\epsfig{figure=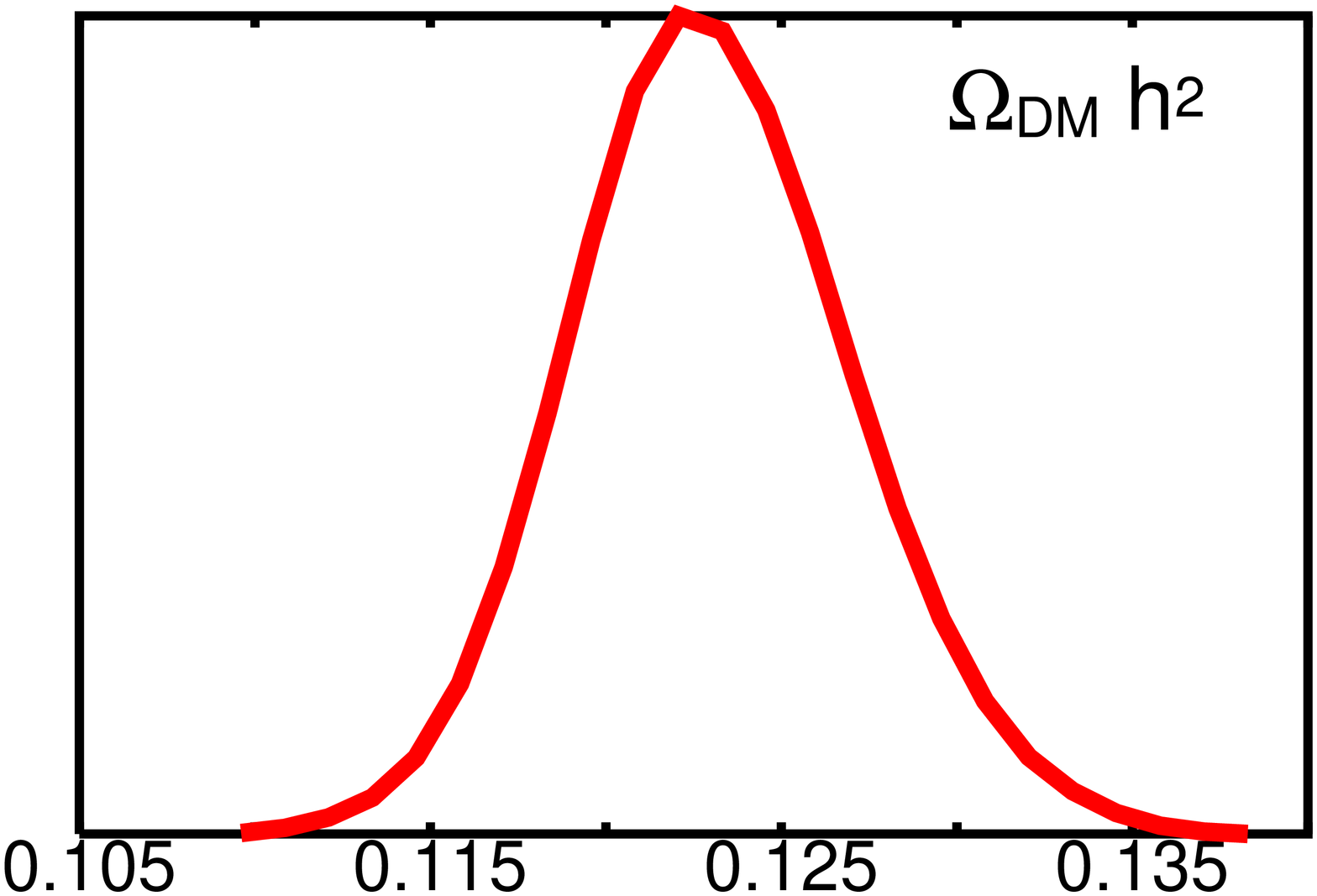,  height=3.6cm, width=3cm, angle=-90 }   & 
\hspace{.5cm} 
\epsfig{figure=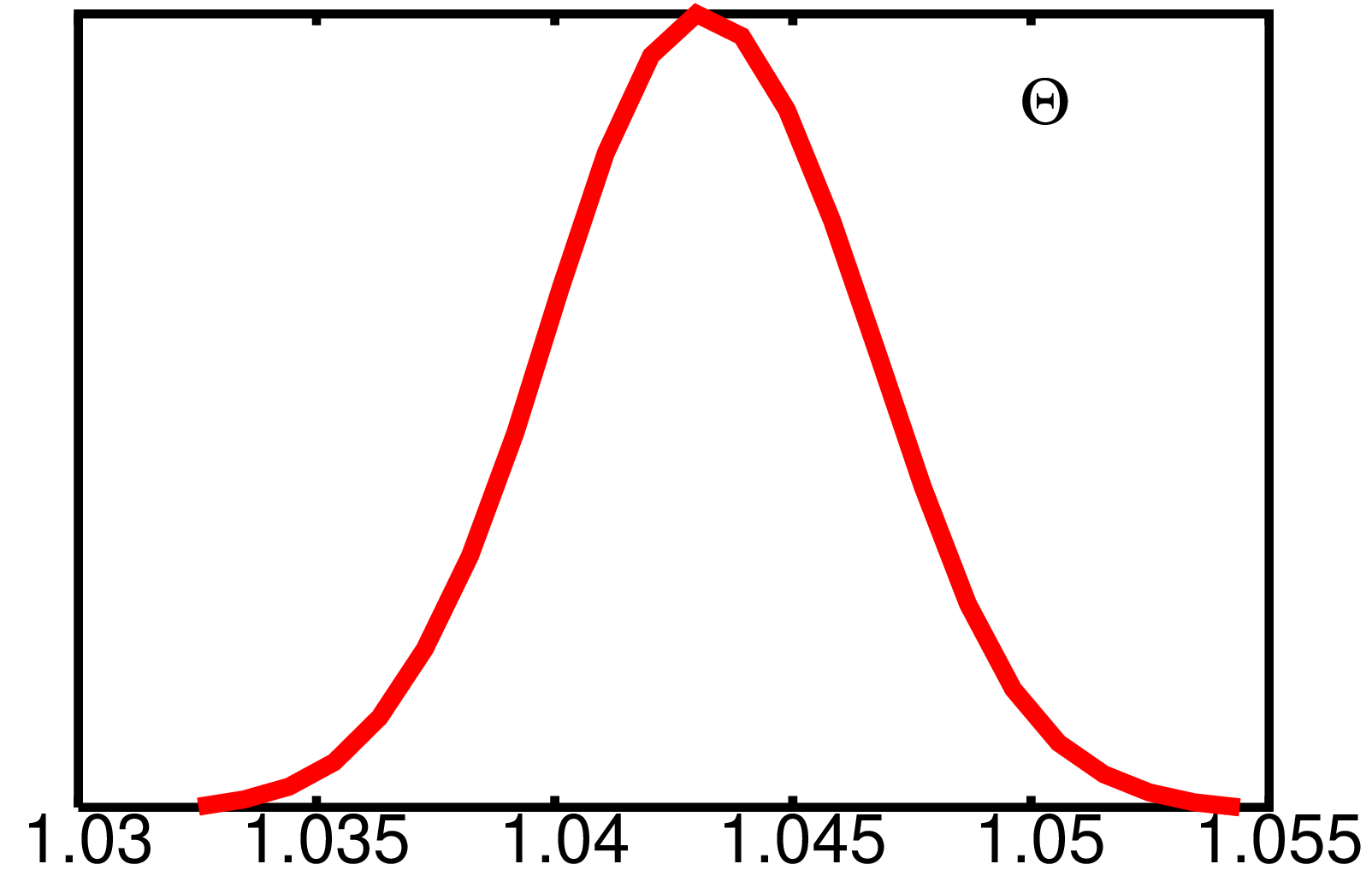,  height=3.6cm, width=3cm, angle=-90 }   \\ 
\hspace{-1cm} 
\epsfig{figure=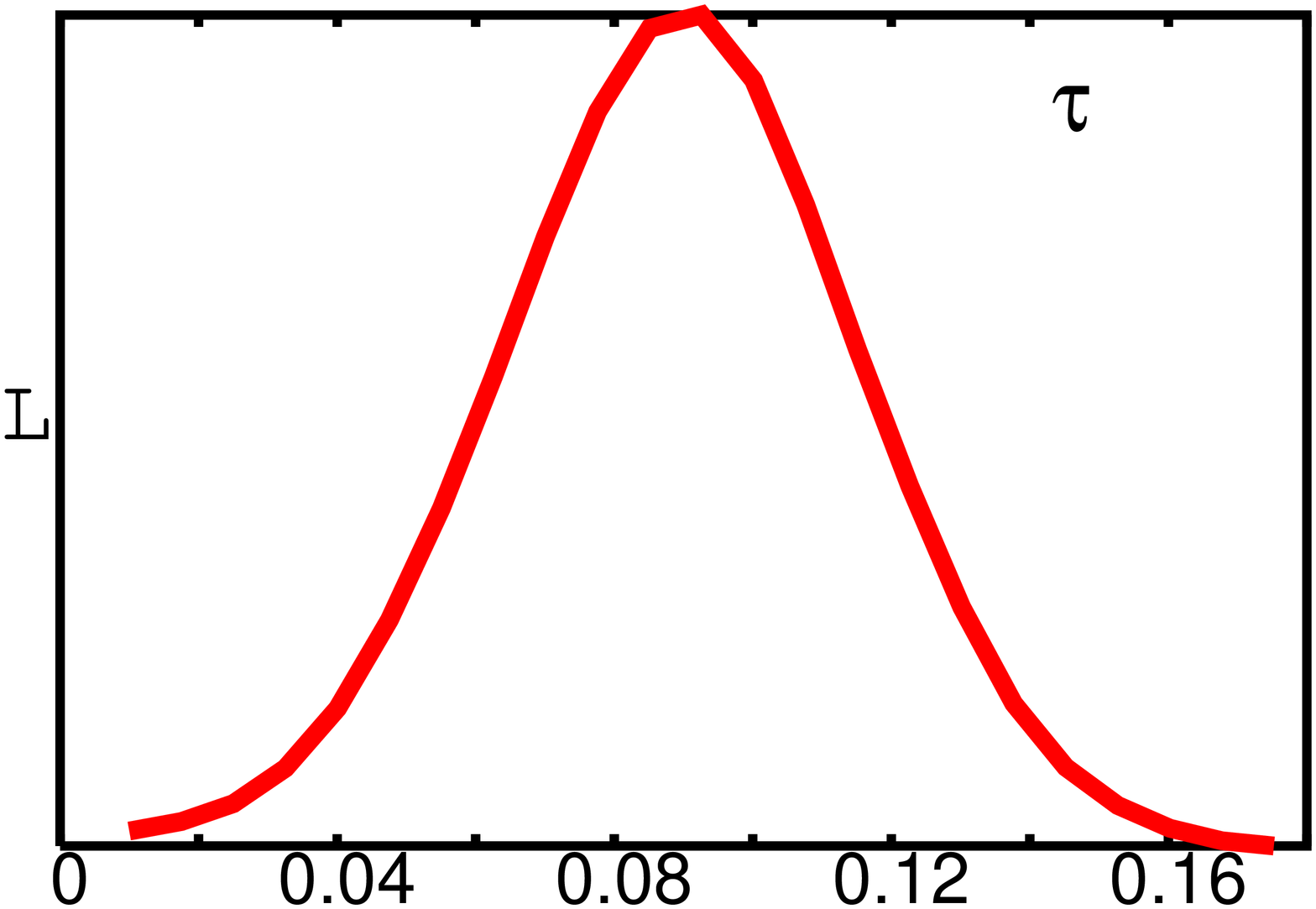,  height=3.6cm, width=3cm, angle=-90 }   & 
\hspace{.5cm} 
\epsfig{figure=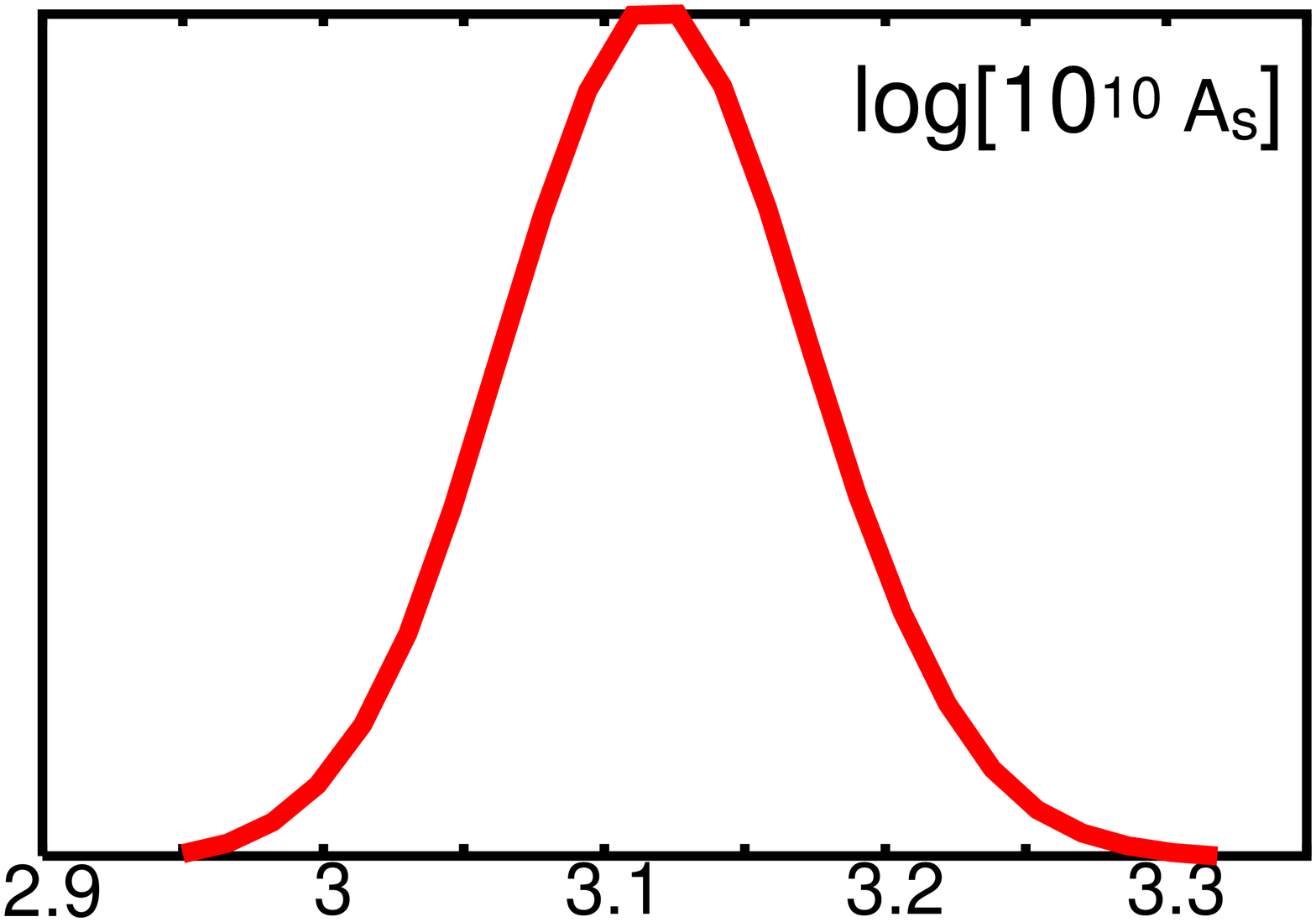,  height=3.6cm, width=3cm, angle=-90 }   & 
\hspace{.5cm} 
\epsfig{figure=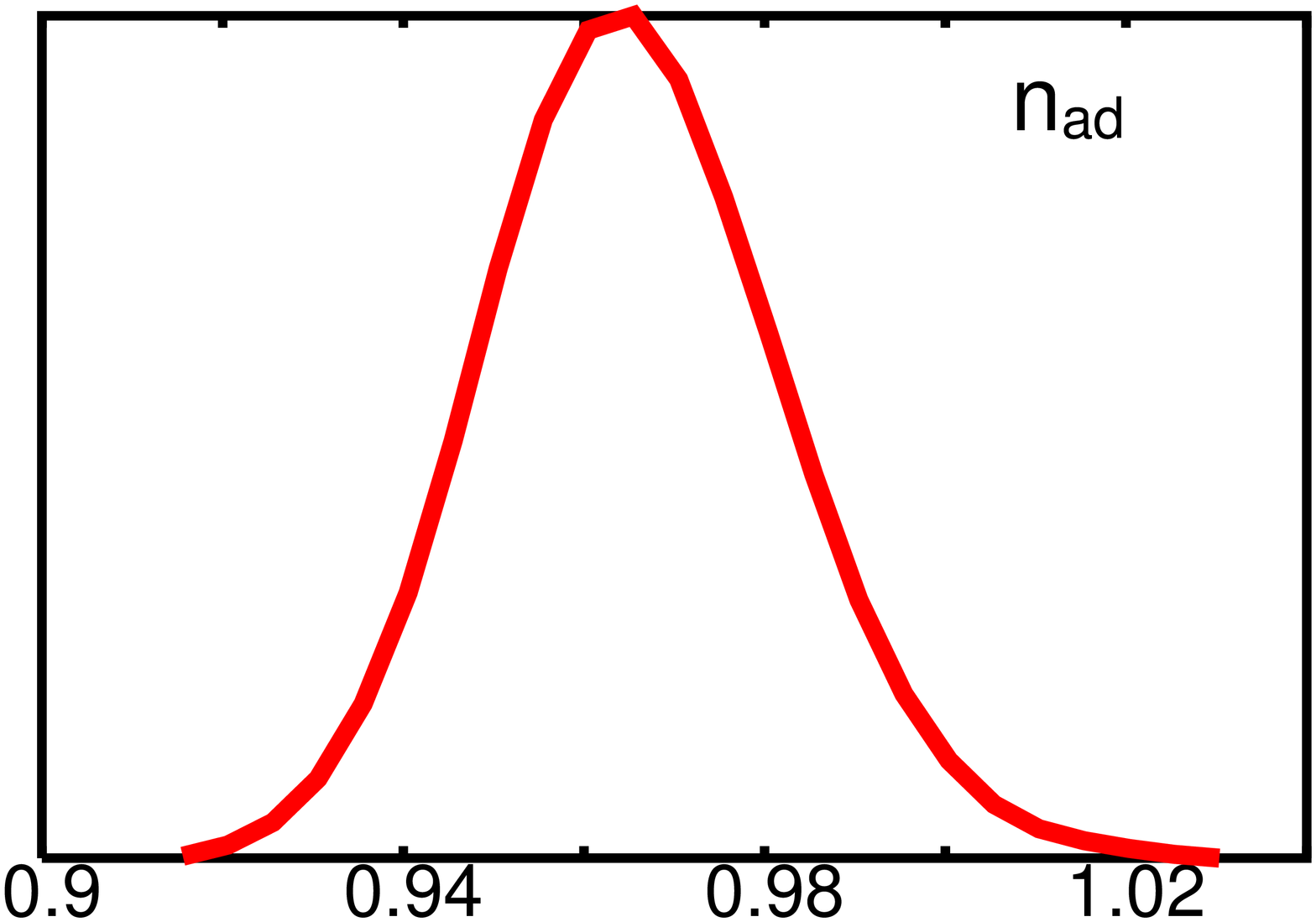,  height=3.6cm, width=3cm, angle=-90 }   &
\hspace{.5cm} 
\epsfig{figure=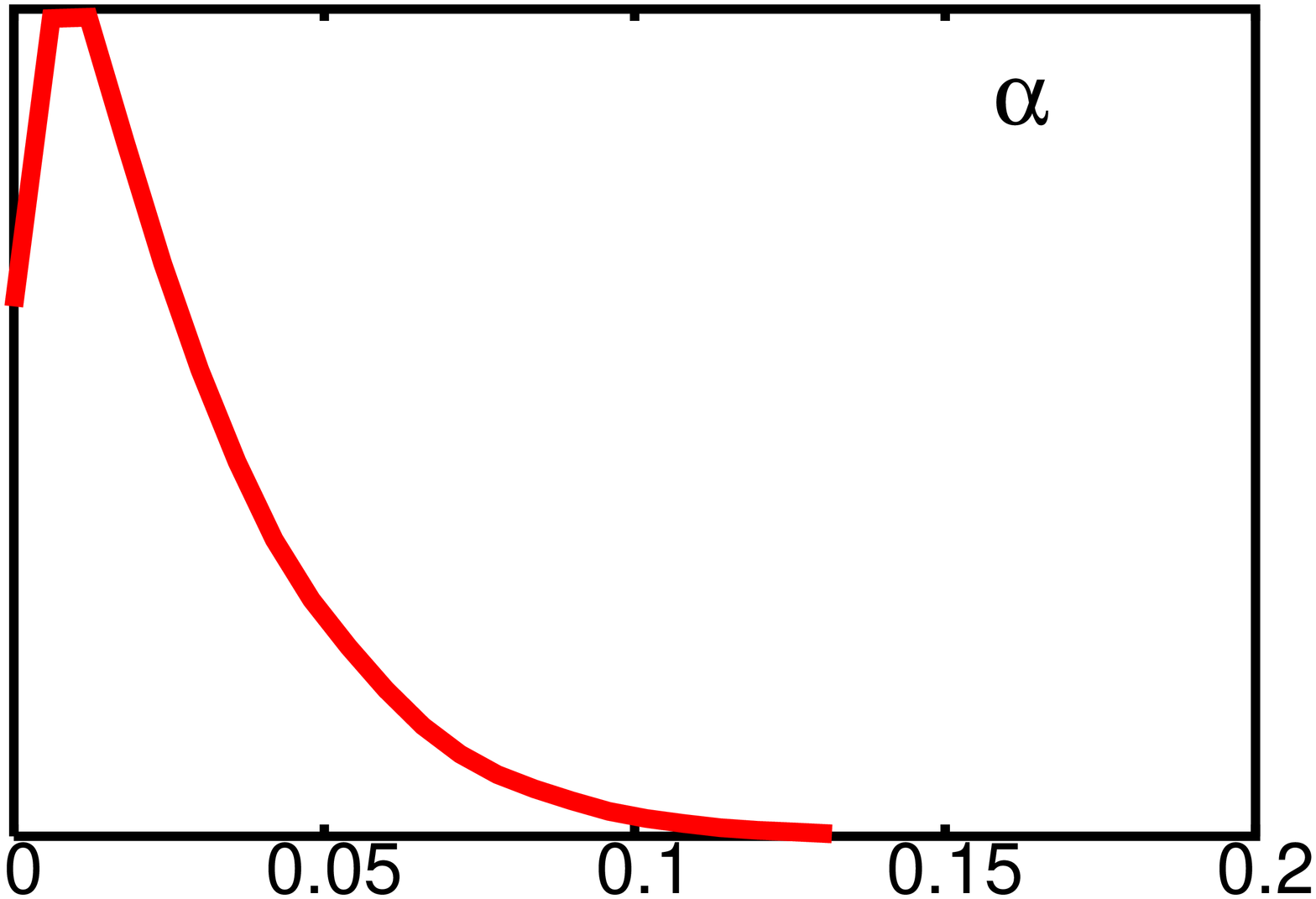,  height=3.6cm, width=3cm, angle=-90 }   
\vspace{0.4cm} 
\end{tabular} 
\caption{ The one dimensional distributions of the sampled parameters. 
} 
\label{axionlikelihood} 
\end{center} 
\end{figure*} 

The one dimensional posterior probability distributions for sampled 
and derived parameters are depicted in Fig.~\ref{axionlikelihood}. 
In particular, the bestfit value for $\alpha$ is  
$\alpha = 8\times10^{-4}$, and the $2\sigma$ bound for the  
marginalized distribution is:  
\be 
\label{alpha_bound} 
\alpha<0.08~\textrm{at}~95\%~\rm{c.l} 
\ee 
while for $\omega_{\rm cdm}$ we get: 
\be 
\label{ocdm_bound} 
\omega_{\rm cdm} =0.123\pm 0.008~\textrm{at}~95\%~\rm{c.l}.  
\ee
Finally, our results for the amplitude of the primordial curvature
spectrum gives the relation \be \epsilon_k \sim 3 \times 10^8 \left(
\frac{M_{\rm inf}}{M_{\rm P}} \right)^4~.  \ee Substituting in
Eq.~(\ref{alpha2}), we see that the bound $\alpha <0.08$ finally
provides the constraint \be \label{final_iso} M_{\rm inf} \leq 10^{13}
{\rm GeV} \left( \frac{f_a}{10^{12}\,{\rm GeV}} \right)^{5/24}~, \ee
corresponding to the central (yellow-)\-shaded forbidden region in
Figs.~\ref{fig:bounds}, \ref{fig:bounds2}. As explained in Section
\ref{isocurvature}, this bound does not apply when thermal
fluctuations induce PQ symmetry restoration after reheating, and we
also assume --as in the rest of the literature-- that it does not hold
when quantum de Sitter fluctuations induce PQ symmetry restoration
during inflation (although we believe that this issue is not
completely clear and deserves further study).  So, for sufficiently
large $H_{\rm inf}$ and/or $\epsilon_{\rm eff}$, the isocurvature
constraint does not apply above a given line in the $(M_{\rm
inf},\,f_a)$ plane, and the allowed region is split in two parts (as
in Fig.~\ref{fig:bounds2}).

\section{Discussion}
\label{discussion}

\subsection{General case}
\label{model-independent}

The constraints discussed in the last section --and in particular
the isocurvature mode limit-- exclude a large region in parameter
space, and {\em preserve only two regions}. 
The first one depends on the reheating 
efficiency\footnote{Here, we are still assuming for simplicity that
$\lambda\sim0.1$. If this is not the case, the
correct result is obtained by replacing the factor
$\epsilon_{\rm eff}$ by $\epsilon_{\rm eff}/(12 \lambda)$.},
\begin{eqnarray}
10^{10} \,{\rm GeV} < f_a < 1.2 \times 10^{11}\,{\rm GeV}\,,
\nonumber \\
5\times 10^{14} F(\epsilon_{\rm eff}, f_a)
\, {\rm GeV}
< M_{\rm inf} < 3 \times 10^{16} \,{\rm GeV}\,, 
\end{eqnarray}
with
\begin{equation}
F(\epsilon_{\rm eff}, f_a) \equiv \min \left\{ 
\sqrt{\frac{f_a}{10^{10} \, {\rm GeV}}} \,\, , \,\, 
\sqrt{\frac{1.4 \times 10^{-8}}{\epsilon_{\rm eff}}}
\frac{f_a}{10^{10} \, {\rm GeV}} \right\} \,.
\end{equation}
The second one, for which the isocurvature mode is too small to 
be excluded by current cosmological data, corresponds to
\begin{equation}\label{strong_minf_bound}
10^{10}\, {\rm GeV} < f_a < 2.5 \times 10^{11}\, {\rm GeV} \,,\hspace{5mm}
M_{\rm inf} < 8 \times 10^{12}\, {\rm GeV}\,,
\end{equation} 
where we assumed that the average misalignment angle in the observable
universe is of order one: otherwise, the upper bound on $f_a$ could be
relaxed significantly, while that on $M_{\rm inf}$ would only increase
slightly, as $f_a$ to the power $5/24$, see Eq.~(\ref{final_iso}).

For a large class of well-known inflationary models, $M_{\rm inf}$ is
typically of the order of $10^{15}$ or $10^{16}$~GeV, and the relevant
allowed window (if any) is the first one. In the next
section, we feature the example of chaotic inflation with a quadratic
potential. 
However, the second window 
(\ref{strong_minf_bound}) is also relevant,
since it is possible to build low-scale inflationary model compatible
with WMAP constraints and satisfying $M_{\rm inf} < 8 \times
10^{12}$~GeV.  Actually, the main motivation for the original hybrid
inflation model of Ref.~\cite{hybrid} was precisely the possibility
to evade axionic isocurvature constraints.  Nowadays, many low scale
inflation models can be built in the generic framework of hybrid
inflation (see Ref.~\cite{LythRiotto} for a review, or~\cite{Allahverdi:2006iq} for
a recent example). In this case,
the isocurvature perturbations carried by the axion are too small
to be constrained by current data, and under the conditions of
(\ref{strong_minf_bound}) the PQ axion model is perfectly viable.

\subsection{Bounds for chaotic inflation with a quadratic potential}
 
Inflation with a 
monomial potential $V(\phi) \propto \phi^{\alpha}$ is usually 
called chaotic inflation. The 
latest WMAP results combined with other data sets essentially rule out 
cases with $\alpha>4$, while the quartic case $\alpha=4$ is only in 
marginal agreement with the data. Therefore, in this section we only 
consider the case of a quadratic potential 
$V(\phi)=\frac{1}{2}m^2\phi^2$, still favored by 
observational bounds. 
 
Using the COBE normalization, it is possible to prove that the mass
should be of order $m \sim 1.3 \times 10^{-6}\,M_{\rm P}$, and to show
that $N_e$ e-folds before the end of inflation,
\be 
\phi_e^2 =  \frac{N_e}{2 \pi} M_{\rm P}^2~. 
\ee 
In particular, between $N_{\rm obs}$ and $N_{\rm QCD}$, the scale of
inflation should be in the range 
\begin{eqnarray}\label{chao-scale}
1.5\times10^{16}\,{\rm GeV} < & M_{\rm inf} & < 
2\times 10^{16}\,{\rm GeV}\,,\nonumber \\
8\times10^{13} \,{\rm GeV} < & H_{\rm inf} & < 
1\times10^{14} \,{\rm GeV}\,.
\end{eqnarray}
So, if $f_a$ is close to $10^{10}$ or $10^{11}$~GeV, one has $H_{\rm
inf}/ 2 \pi \geq f_a$ and the PQ symmetry is broken by quantum
fluctuations during inflation. If these fluctuations erase the
isocurvature modes on cosmological scales as suggested by
Ref.~\cite{LythStewart2}, then chaotic inflation is compatible with
the axion scenario.
  
\begin{figure}
\begin{center}
\begin{tabular}{ll} 
\psfig{figure=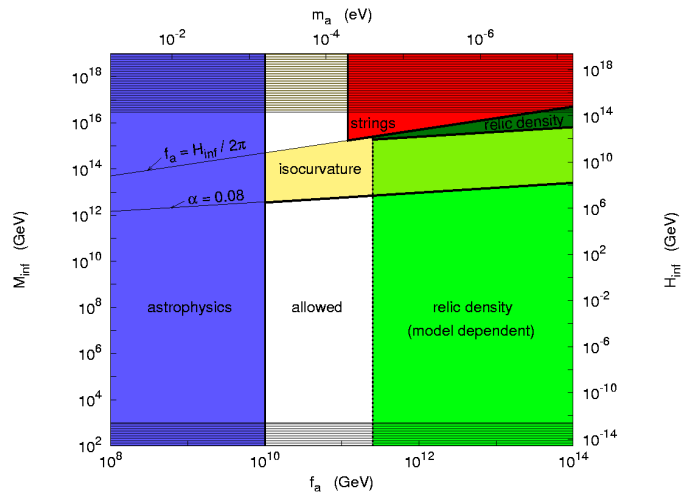,width=0.5\textwidth} 
\end{tabular} 
\caption{ Bounds in the $(M_{\rm inf},\,f_a)$ plane, assuming that
  reheating is very inefficient and the PQ symmetry can never be
  thermally restored ($\epsilon_{\rm eff} \leq 4 \times 10^{-12}$).
  The energy scale of inflation $M_{\rm inf}$ is bounded from above by
  $r<0.3$ (top hatched) and from below by the requirement of
  successful baryogenesis (bottom hatched).  Possible values of the PQ
  scale $f_a$ lie between the regions excluded by supernovae (left
  blue) and by the bound on the axionic relic density, produced by one
  of the following two mechanisms: axionic string decay (red) or
  misalignment angle at the time of the QCD transition (green).  For
  the bounds related to the relic density, we use a solid line for the
  model independent bound (with a minimal, fine-tuned value of the
  misalignment angle), and a dotted line for the model dependent bound
  (with a generic misalignment angle of order one).  The yellow region
  is excluded by the limit on the isocurvature mode amplitude
  $\alpha$: this is the main result of this work. If large quantum
  fluctuations during inflation can erase isocurvature modes on
  cosmological scale (as suggested by Ref.~\cite{LythStewart2}), the
  isocurvature bound does not apply above the line corresponding to
  $f_a = H_{\rm inf}/2 \pi$.  The remaining allowed regions are left
  in white. Details are explained in the text.}
\label{fig:bounds} 
\end{center}
\end{figure}

\begin{figure}
\begin{center}
\begin{tabular}{ll} 
\psfig{figure=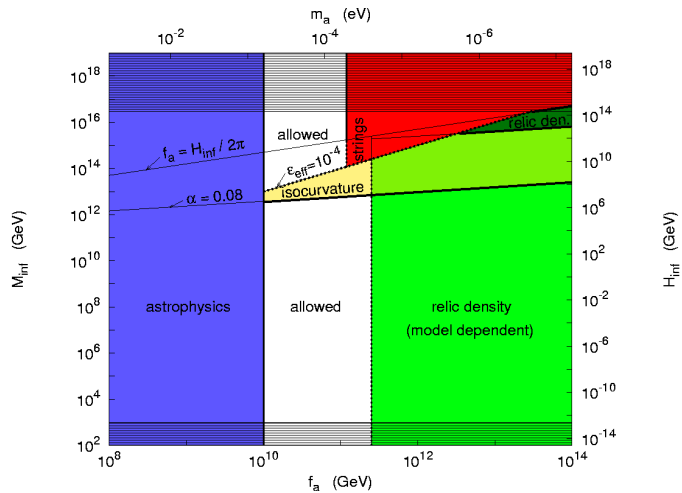,width=0.5\textwidth} 
\end{tabular} 
\caption{ Same as previous figure, but assuming now 
  that reheating is very efficient ($\epsilon_{\rm eff} =
  10^{-4}$) so that the the PQ symmetry is thermally restored above
  the dotted line.  In this case, the constraint from string
  production and decay (red) extends to smaller values of $M_{\rm
    inf}$, but above the dotted line and for $10^{10} {\rm GeV} < f_a <
  1.2 \times 10^{11}{\rm GeV}$ there is a new allowed region
  (corresponding to the erasure of isocurvature perturbations in
  the thermal bath).}
\label{fig:bounds2} 
\end{center}
\end{figure}

We could hope to find a second allowed window in the case where
$f_a$ is very large and the misalignment angle is fine-tuned to
a very small value (i.e., inside the light green region in 
Figs.~\ref{fig:bounds}, \ref{fig:bounds2}). Note that this is not
possible, at least in the case of chaotic inflation.
Indeed, reheating after chaotic inflation is expected to lead to
a temperature $T_{\rm rh}$ of the order of $10^{10}$ or $10^{11}$~GeV.
So, if $f_a$ is much bigger than $10^{11}$GeV, thermal symmetry 
restoration after inflation cannot take place, and isocurvature bounds
are applicable. Given the value of $M_{\rm inf}$ in Eq.~(\ref{chao-scale}),
the isocurvature constraint (\ref{final_iso}) would enforce 
$f_a \geq 7\times10^{27} \,
{\rm GeV} \gg M_{\rm P}$, which is not realistic.
 
\subsection{Possible loopholes} 
 
In this subsection we will explore those loopholes we have left 
open for the axion to be the dominant component of cold dark matter. 
 
\subsubsection{Production of cosmic strings during reheating} 
 
Even if the reheating temperature of the universe is too low for a 
thermal phase transition at the Peccei-Quinn scale, it is possible 
that axionic cosmic strings be formed during preheating if the field 
responsible for symmetry breaking at the end of inflation is the 
Peccei-Quinn field. Then, the residual global $U(1)$ symmetry of the 
vacuum gives rise to axionic cosmic strings. At present there is no 
prediction for what is the scaling limit of such a mesh of strings 
produced during preheating. A crucial quantity that requires 
evaluation is the fraction of energy density in infinite strings 
produced at preheating, since they are the ones that will give the 
largest contribution to the axion energy density. It could then be 
that axionic strings produced at preheating may be responsible for the 
present axion abundance. We leave for the future the investigation of 
this interesting possibility.

\subsubsection{Axion dilution by late inflation} 
 
When imposing bounds on the axion mass from its present abundance, it
is assumed that no there is no significant late entropy
production~\cite{dilution} or dilution from a secondary stage of
inflation. We truly don't know.  It has been speculated that a short
period of inflation may be required for electroweak baryogenesis to
proceed~\cite{Garcia-Bellido:1999}.  In such a case, a few e-folds
($N\sim5$) of late inflation may dilute the actual axion energy
density by a factor $\gamma = e^{-3N}\sim10^{-7}$, easily evading the
bounds.

\subsubsection{Coupling between radial part of PQ field and other fields.} 

The radial part of the PQ field could interact with other fields,
including the inflaton (see e.g.~\cite{hybrid}). Then the effective
$f_a$ during inflation might be anything, and the isocurvature
bounds might be alleviated.
 
\section{Conclusions}\label{conclusions} 
 
In this paper, we reanalyzed the bounds on the cosmological scenario
in which the cold dark matter is composed of axions plus some other
component (like e.g. neutralinos), and the energy density of axions is
produced by the misalignment mechanism at the time of the QCD
transition. In this case, a fraction of the cosmological
perturbations consists of isocurvature modes related to the quantum
perturbations of the axion during inflation. We use this possible
signature plus the contribution to dark matter energy density as a
tracer of axions in the universe.  In that sense, this work is similar
to that presented in \cite{Sikivie:2006ir}. However, we make a
stronger statement about the generation of isocurvature modes in the
case in which the PQ symmetry is restored by quantum fluctuations
during inflation, and we significantly improve previous bounds coming
from the non-observation of isocurvature modes in the CMB, in
particular, given the recent WMAP 3-year data.

A narrow window for the PQ scale $10^{10}\,{\rm GeV} < f_a <
10^{12}$~GeV still remains open, but we show that the main consequence
of recent data on cosmological perturbations is to limit the possible
energy scale of inflation in this context, in order not to have an
excessively large contribution of isocurvature perturbations to the
CMB anisotropies. In particular, we show that in the axion scenario,
the energy scale of inflation cannot be in the range $8 \times 10^{12}
< M_{\inf} < 5 \times 10^{14}$~GeV, unless one of the two situations
occurs: either reheating leads to a temperature $T_{\rm rh}>f_a$, with
an efficiency parameter $\epsilon_{\rm eff} \geq 4 \times 10^{-12}$,
and there is another allowed region with $6 \times 10^{10}
\epsilon_{\rm eff}^{-1/2} \,{\rm GeV} < M_{\rm inf} < 3 \times
10^{16}$~GeV; or the misalignment angle is fine-tuned to very small
values (this coincidence can be motivated by anthropic considerations)
and the upper bound $M_{\inf} < 10^{13}$~GeV can be slightly weakened
(by at most one order of magnitude).

These bounds may be of interest taking into account particle physics
experiments searching for the axion, which may help to put very
stringent constraints on inflationary models.  Detecting the QCD axion
could shed some light on the scale of inflation and possibly into the
mechanism responsible for inflation.

In this respect, there is an intriguing possibility that the PVLAS
experiment~\cite{PVLAS} may have observed a pseudo-scalar particle
coupled to photons.  The nature of this particle is yet to be decided,
since its properties seem in conflict with present bounds on the axion
coupling to matter~\cite{ADMX,CAST}, see however~\cite{Masso,Mendonca}.

\section*{Acknowledgements.}

It is a pleasure to thank Andrei Linde, David Lyth, Pierre Sikivie and
David Wands for enlightening discussions.  We also thank A. Lewis for
providing the updated CosmoMC code with the WMAP 3-year data. JGB and
JL thank the Galileo Galilei Institute for Theoretical Physics and
INFN for hospitality and support during the completion of part of this
work. We acknowledge the use of the MareNostrum Supercomputer at the
Barcelona Supercomputing Center.  This work was supported in part by a
CICYT project FPA2003-04597.  The work of MB was also supported by the
``Consejer\'\i a de Educaci\'on de la Comunidad de Madrid--FPI
program''. We acknowledge the financial support of the CICYT-IN2P3
agreement FPA2005-9 and FPA2006-05807.

\end{document}